\documentclass[twocolumn,preprintnumbers,amsmath,amssymb]{revtex4}
\usepackage{graphicx}% Include figure files
\usepackage{dcolumn}% Align table columns on decimal point
\usepackage{bm}% bold math

\newcommand{\comment}[1]{}
\newcommand{\rb}{$^{87}$Rb }

\newcommand{\Da}{\Delta_{ca}}
\newcommand{\DN}{\Delta_N}
\newcommand{\Dc}{\Delta_{pc}}
\newcommand{\omz}{\omega_z}
\newcommand{\omrec}{\omega_{\rm rec}}
\newcommand{\nb}{\bar{n}}

\newcommand{\nbar}{\bar{n}}

\newcommand{\nbarnl}{\bar{n}_{\rm nl}}
\newcommand{\gamosc}{\Gamma_{\rm osc}}

\usepackage{color}

\begin{document}

\title{Cavity Nonlinear Optics at Low Photon Numbers from Collective Atomic Motion}

\author{Subhadeep Gupta$^{1}$}
\author{Kevin L.\ Moore$^{1}$}
\author{Kater W.\ Murch$^{1}$}
\author{Dan M.\ Stamper-Kurn$^{1,2}$}
\affiliation{
    $^1$Department of Physics, University of California, Berkeley CA 94720 \\
    $^2$Materials Sciences Division, Lawrence Berkeley National Laboratory, Berkeley, CA 94720}
\date{\today}% It is always \today, today,
             % but any date may be explicitly specified

\begin{abstract}

We report on Kerr nonlinearity and dispersive optical bistability
of a Fabry-Perot optical resonator due to the displacement of
ultracold atoms trapped within. In the driven resonator, such
collective motion is induced by optical forces acting upon up to
$10^5$ $^{87}$Rb atoms prepared in the lowest band of a
one-dimensional intracavity optical lattice. The longevity of
atomic motional coherence allows for strongly nonlinear optics at
extremely low cavity photon numbers, as demonstrated by the
observation of both branches of optical bistability at photon
numbers below unity.

\end{abstract}

%\pacs{03.75.Gg,05.30.Jp,52.38.Bv}

\maketitle

Nonlinear optical phenomena occur typically at high optical
intensities, or, equivalently, at high average photon numbers
$\bar{n}$ in an optical resonator, because conventional materials
mediate weak coupling between photons.  Producing such phenomena
at very low photon numbers, i.e.\ $\bar{n} \leq 1$, is desirable
for applications ranging from optical communication to quantum
computation \cite{cira97trans,imam97strong}. In this low-intensity
regime, nonlinear optics requires the use of materials with
optical properties that are significantly altered even by single
photons, and in which this alteration persists long enough to
impact the behavior of subsequent photons interacting with the
material.

Such requirements may be satisfied by atoms within a high-finesse,
small-volume optical resonator. Under the condition of collective
strong coupling, defined as $N C = N (g^2 / 2 \kappa \Gamma) \gg
1$, atomic saturation on an optical transition induces nonlinear
effects such as absorptive optical bistability
\cite{remp91bistab,grip96bistab,saue04cqed} with $N \gg 1$, or
cross phase modulation \cite{turc95phase} and photon blockade
\cite{birn05block} with $N \simeq 1$. Here, $N$ is the number of
atoms in the resonator, $g$ the atom-cavity coupling frequency,
and $\kappa$ and $\Gamma$ the cavity and the atomic coherence
decay rates, respectively. Since strong nonlinearities (e.g.\ the
upper branch of the optical bistability curve) occur beginning at
$\nbarnl \simeq N (\Gamma / \kappa)$, the condition $\nbarnl \leq
1$ requires that the atomic coherence persist longer than the
residence time of photons in the cavity. This condition is met
with neutral atoms in state-of-the-art Fabry-Perot cavities
\cite{turc95phase,birn05block} and in recent experiments with
superconducting circuits within microwave resonators
\cite{wall04cqed}. Strong nonlinearities are also achieved using
the long coherence times for atomic Raman transitions
\cite{merr00,hemm95,wang01}. The lowest reported resonator photon
number for observing both branches of bistability in an optical
cavity is $\nbarnl \simeq 100$ \cite{remp91bistab,saue04cqed}.

The motional degrees of freedom of ultracold atomic gases represent
a new source of long-lived atomic coherence affecting light-atom
interactions. This coherence leads, for example, to superradiant
light scattering at low threshold intensities both in free space
\cite{inou99super2,yosh05super} and inside optical resonators
\cite{slam07super}.  The motion of single atoms in strong-coupling
cavities has been observed and cooled,
\cite{hood00micro,pink00trap}, and effects of motion in atomic
ensembles in weak-coupling cavities have been examined
\cite{chan03,krus03ring,nago03coll,klin06nms,slam07super}.

We report here on nonlinear optics arising from long-lived
coherent motion of ultracold atoms trapped within a high-finesse
Fabry-Perot cavity. Optical forces exerted by light within the
cavity displace the trapped atoms so as to significantly vary
their position-dependent coupling to the cavity mode.  The
consequent refractive nonlinearity and bistability are observed in
the transmission of probe light through the cavity. Because the
coherence time (1 ms) for atomic motion is much longer than the
residence time of photons in the cavity ($1 / 2 \kappa = 120$ ns),
strong optical nonlinearity is observed even at $\nbar = 0.05$,
and is predicted to occur as low as $\nbar \simeq 10^{-4}$.

To explain this nonlinear optical behavior, we consider the
one-dimensional motion of $N$ atoms in a Fabry-Perot cavity in
which the atom-cavity coupling frequency varies along the cavity
axis as $g(z) = g_0 \sin k_p z$. A harmonic potential of the form
$V(z) = m \omega_z^2 (z - z_0)^2 / 2$ confines the atoms.  For a
trapping frequency $\omega_z \gg \hbar k_p^2 / 2 m$, the position
of atoms in the trap ground state is determined to below the
optical wavelength $2 \pi / k_p$. In the dispersive regime, with
the detuning $\Da = \omega_c - \omega_a$ between the bare-cavity
and atomic resonances being large ($|\Da| \gg \sqrt{N} g_0$), the
atomic ensemble presents a refractive medium in the cavity that
shifts the cavity resonance by $\DN \simeq N g(z_0)^2/\Da$.

\begin{figure}[b]
\includegraphics[angle = 0, width = 0.5 \textwidth] {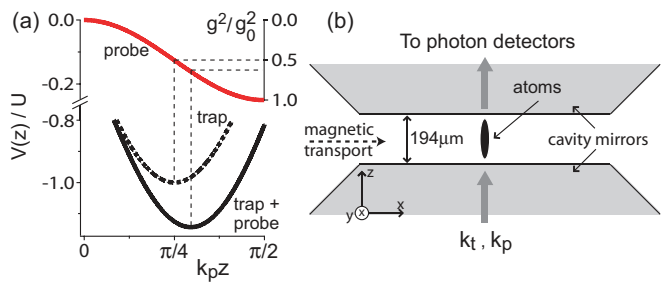}
\caption{\label{setup} (a) The optical potential due to either the
cavity probe light (red, here with maximum depth $U/4$ and
$\Da<0$), or the trapping light (black dashed, with maximum depth
$U$) is shown near the axial minimum of one lattice site. The trap
minimum of the combined optical potential (black, solid) is
displaced in the presence of the probe, changing the atom-cavity
coupling strength ($\propto g(z)^2$), and, thereby, the cavity
resonance frequency. (b) Experimental setup: ultracold \rb atoms
are trapped in a standing wave of 850 nm light formed by exciting
a TEM$_{00}$ mode of a vertically oriented Fabry-Perot cavity. The
system is probed by detecting transmission through another
TEM$_{00}$ mode at $k_p = 2 \pi/ 780$ nm.} \label{fig:setup}
\end{figure}

Probe light in the cavity produces an additional, optical force on
the atoms with average strength $f(z_0) \nbar$ where $f(z) = - \hbar
\partial_z g^2(z) / \Da$. This force displaces the equilibrium
position of the atoms by $(f(z_0) / m \omz^2) \nbar$. In turn, this
displacement (assumed small) varies the atom-cavity coupling
strength (Fig.\ \ref{fig:setup}(a)). Taking $k_p z_0 = \pi/4$ for
illustrative purposes, the cavity resonance shift changes to $\DN(1-
\epsilon \nbar)$ with $\epsilon = 2 \hbar k_p^2 g_0^2 / m \Da
\omz^2$. Thus, this system is characterized by a Kerr nonlinearity,
with the refractive index of the intracavity medium, $1 +
(\DN/\omega_p)( 1 - \epsilon \nb) = n_0 + n_2 I_p$ varying with the
probe intensity $I_p \propto \nb$.

This nonlinearity affects the frequency response of the cavity as
evident, for example, in the intracavity photon number
$\nbar(\Dc)$ for probe light detuned by $\Dc$ from the bare cavity
resonance. Assuming the atoms adiabatically follow the trap
minimum, one obtains
\begin{equation}
\bar{n}(\Dc) = \frac{\bar{n}_{\rm max}}{1+\left(\frac{\Dc - \DN
(1-\epsilon \bar{n})}{\kappa}\right)^2} \label{eq:lineshape}
\end{equation}
where $\bar{n}_{\rm max}$ defines the probe intensity. This
lineshape is equivalent to that obtained for a damped nonlinear
oscillator \cite{land76mech}. The deviation from the Lorentzian
lineshape of the bare cavity may be quantified by the parameter
$\beta$ that gives the maximum nonlinear shift of the cavity
resonance in units of the cavity half-linewidth. Here $\beta =
(\DN \epsilon / \kappa) \nb_{\rm max}$. In particular, for $\beta
> 8 \sqrt{3}/9 \simeq 1.54$, the cavity exhibits refractive
bistability \cite{boyd03}.

In our experiment (Fig.\ \ref{fig:setup}(b)), atoms were trapped not
in a single harmonic trap, but rather in a one-dimensional optical
lattice comprising many potential wells. This optical trap was
formed by coupling frequency-stabilized laser light with a
wavenumber of $k_t = 2 \pi / 850$ nm to a TEM$_{00}$ mode of a
Fabry-Perot cavity that was actively stabilized with this same
light. The trap was loaded with an ultracold $^{87}$Rb gas in the
$|F=1, m_F = -1\rangle$ hyperfine ground state that was transported
magnetically to within the optical cavity prior to being transferred
to the optical trap. Evaporative cooling in the optical trap yielded
a gas of up to $10^5$ atoms occupying $\simeq 300$ adjacent lattice
sites at a temperature of $T = 0.8 \, \mu$K. The trap depth was set
to $U \simeq k_B \times 6.6 \, \mu$K, corresponding to axial and
radial trapping frequencies of $\{\omz, \omega_{x,y}\} \simeq 2 \pi
\! \times \! \{42, 0.3\}$ kHz, about each minimum of the optical
trap. Given $\hbar \omz / k_B = 2 \, \mu \mbox{K} > T$, the gas
predominantly occupied the ground state for axial vibration in each
lattice site.

According to the specific value of $k_t$, another TEM$_{00}$ mode
of the cavity was stabilized with $|\Da| = 2 \pi \! \times \! (10
- 300)$ GHz detuning from the D2 resonance line at a wavenumber of
$k_p = 2 \pi / 780$ nm.  The cavity finesse for this mode was
measured as ${\mathcal{F}_{780}} = 5.8 \times 10^5$, corresponding
to a resonance half-linewidth of $\kappa = 2 \pi \! \times \!
0.66$ MHz given the 194 $\mu$m spacing between the cavity mirrors.
From the cavity mode waist of $w_0 = 23.4\, \mu$m and summing over
all states excited by the $\sigma^+$ polarized light used in our
experiment, we determine an atom-cavity coupling frequency of $g_0
= 2 \pi \! \times \! 14.4$ MHz at the antinodes of the cavity
field. Given the atomic half-linewidth of $\Gamma = 2 \pi \!
\times \! 3$ MHz, this cavity setup achieves the single-atom
strong coupling conditions, with critical atom ($2 \Gamma \kappa /
g_0^2 = 0.02$) and photon ($\Gamma^2 / 2 g_0^2 = 0.02$) numbers
below unity.

\begin{figure}[b]
\includegraphics[angle = 0, width = 0.5 \textwidth] {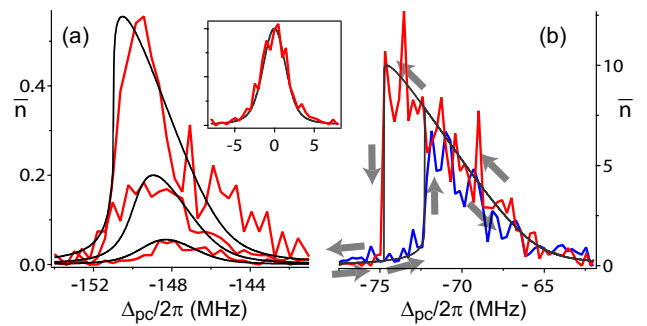}
\caption{Nonlinear and bistable cavity lineshapes shown as the
intracavity photon number $\nbar$ vs.\ detuning $\Dc$ from the
bare-cavity at constant input power. (a) The cavity lineshape (red,
single (highest $\nb_{\rm max}$) or average of 5 (others)
measurements) becomes increasingly asymmetric as the input intensity
is increased. The triggering technique (see Fig.\ \ref{ring}(a))
used to obtain these spectra resulted in variations of up to
$1\,$MHz in $\DN/2 \pi$ across measurements. Model lineshapes
(black) using the Voigt profile indicated by the bare-cavity
lineshape (inset) are calculated with $\beta=\{0.37,1.33,3.72\}$
determined from the observed $\nbar_{\rm max}=\{0.06,0.20,0.56\}$, a
common $\Delta_N = -2 \pi \! \times \! 148$\,MHz, and the known $\Da
= -2 \pi \! \times \! 30$\,GHz, and $\omz = 2 \pi \! \times \!
42\,$kHz. (b) The lower (blue) and upper (red) branches of optical
bistability are observed in a single experiment with probe light
swept with opposite chirps ($\pm 6$ MHz/ms) across the cavity
resonance, and match the expected behavior (black) with $\beta=9.5$,
$\Da = - 2 \pi \! \times \! 101\,$GHz, and $\omz = 2 \pi \! \times
\! 42\,$kHz \cite{kerrnote}.} \label{fig:kerrlines}
\end{figure}

After preparing the intracavity atomic medium, its nonlinear optical
response was probed by detecting the transmission of light at
wavenumber $k_p$ through the cavity. For this measurement, the probe
light was switched on and its frequency $\omega_p/2 \pi$ was swept
linearly across the cavity resonance at a rate of a few MHz/ms. The
sweep rate was sufficiently slow so that trapped atoms adiabatically
followed the variation in the intracavity optical potential, and
sufficiently fast so that atom loss from probe-induced heating
\cite{hech98cool} was negligible. The transmitted probe light was
separated from the trapping light using several filters, and was
then detected by single-photon counters. The overall efficiency of
detecting  an intracavity photon was $\simeq 0.05$.

As the power of the probe light was increased, the observed cavity
transmission lineshapes increasingly deviated from the symmetric
lineshapes observed for the bare cavity and at low powers (Fig.\
\ref{fig:kerrlines}(a)). This behavior can be described by adapting
the model developed above to the experimental case with atoms
trapped in a multitude of potential wells. Due to the difference
between the wavelengths of the trapping and the probing light, atoms
in each well of the optical trap experience a different strength and
gradient of the probe-induced AC Stark shift. Nevertheless, at all
potential wells, this Stark shift displaces all atoms so as to
either increase (for $\Da < 0$) or decrease ($\Da > 0$) their
coupling to the cavity. That is, the regular crystalline arrangement
of atoms in the confining optical lattice is distorted by probe
light, somewhat analogously to the mechanism for Kerr effects in
solids. Eq.\ \ref{eq:lineshape} remains valid for our experiment
with $\epsilon$ halved to account for averaging over the many wells.
The contribution of radial motion to this effect is negligible.

To compare these measurements with theoretical predictions, we
also account for technical fluctuations in the probe detuning
$\Dc$. The bare-cavity transmission lineshape (Fig.\
\ref{fig:kerrlines} inset) is well approximated by the convolution
of a Lorentzian with half-linewidth $\kappa$ and a Gaussian with
rms width of $\sigma = 2 \pi \! \times \! 1.1$ MHz. Replacing the
Lorentzian of Eq.\ \ref{eq:lineshape} with this Voigt profile, and
using values of $\beta$ determined from measured experimental
parameters, we obtain excellent agreement with the transmission
measurements (Fig.\ \ref{fig:kerrlines}(a)). The effect of the
internal state dynamics alone in producing nonlinear optical
behavior is negligible for our experimental parameters
\cite{intnote}.

At sufficiently strong refractive nonlinearity, the optical
resonator becomes hysteretic and bistable. To observe clearly this
bistability, we measured the transmission lineshape of the cavity
resonance with consecutive sweeps of the probe frequency with
opposite chirp, i.e.\ sweeping linearly towards and then away from
the bare cavity resonance. For this measurement, the cavity was
operated at $\Da = - 2 \pi \! \times \! 101\,$ GHz and contained
$N \simeq 7\times 10^4$ atoms. For $\nbar_{\rm max} = 10$, the
parameter $\beta = 9.5$ is well within the bistable regime, which,
for the Voigt profile relevant to our system, occurs for $\beta
\gtrsim 3.7$. The observed lineshapes (Fig.
\ref{fig:kerrlines}(b)) exhibit several hallmarks of bistability,
such as the abrupt changes in $\nbar$ at the termini of the upper
and lower branches of stability curves and the difference in the
maximum intracavity photon numbers attained for the upward and
downward frequency sweeps. These features were also observed in
similar experiments with $\Da = - 2 \pi \! \times \! 10$ GHz and
at the lowest intensities detectable in our system, with $\nbar =
0.05$.

In previous studies of bistability in cavity QED, the operating
range over which hysteretic behavior was observed was narrower
than predicted theoretically. In contrast with those studies, here
the cavity state is determined by atomic motion that varies only
at a timescale of $\omz^{-1}$. As this timescale is much longer
than both the $1/\kappa$ timescale for fluctuations of the
intracavity field and also the typical $\sim 10 \, \mu$s timescale
of variations in $\Dc$, such fluctuations do not destabilize the
atoms-cavity system, and the full bistable region predicted by our
theoretical treatment is observed.

As the detuning $\Da$ is reduced, the minimum photon number
$\nbarnl$ for strong nonlinearities
\begin{equation}
\nbarnl = 4 \frac{\omz^2 \kappa}{\omrec \, g_0^2} \frac{ N g_0^2 / 2
+ \left( \Da/2\right)^2}{N g_0^2},
\end{equation}
as defined by the condition $\beta = 1$, diminishes.  Here, $\omrec
= \hbar k_p^2 / 2 m$ is the probe recoil frequency and the effects
of spatial averaging are included.  For $\Da \rightarrow 0$, this
number reaches a limiting value of $\nbarnl \simeq 10^{-4}$ for a
minimum value for the trapping frequency of  $\omz = 2 \omrec$ and
other parameters of our experiment.

\begin{figure}[b]
\includegraphics[angle = 0, width = 0.5 \textwidth] {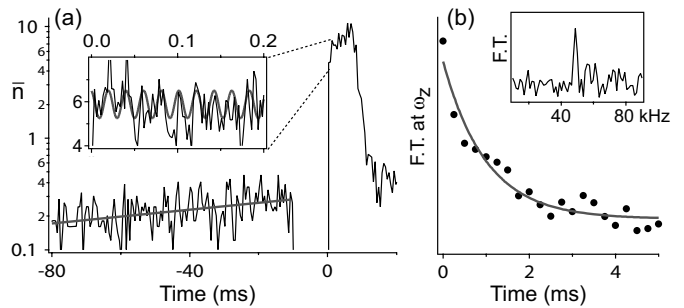}
\caption{\label{ring} Measurement of collective atomic  motion.
(a) A record of the cavity transmission demonstrates the
experimental sequence of a trigger, delay and detection phase.
Inset: coherent atomic motion modulates the cavity transmission at
$\omz=2 \pi \! \times \! 49\,$kHz (shown is the first $200 \,
\mu$s of the detection phase, 50-measurement average, $400\,$kHz
bandwidth), in agreement with the gray theory curve showing the
expected amplitude ($2 \beta = 0.75$) and phase of the transmitted
signal. (b) Decay of the collective atomic motion. The Fourier
spectral amplitude at $\omz$, determined over $500 \, \mu$s
intervals of the transmission measurement, decays within a $1/e$
time of $1.0(1)\,$ms (50-measurement average). Inset: Fourier
amplitude spectrum of the first $500 \, \mu$s of the detection
phase for a typical single measurement.} \label{fig:ring}
\end{figure}

For $\nbarnl < \omz / 2 \kappa \simeq 0.03$, the optical
nonlinearity stems from the passage through the cavity of less
than one photon on average per trap oscillation period. Photon
number fluctuations may then cause a significant non-adiabatic
atomic motional response. Taking the characteristic time for these
fluctuations to be $1 / 2 \kappa$, a single photon imparts an
impulse $f(z) / 2 \kappa$ to an atom at location $z$. This impulse
induces transient motional oscillations in the atomic medium that
modulate the cavity resonance frequency. For our experimental
parameters, with $N = 5 \times 10^4$ atoms, this modulation
becomes greater than the cavity half-linewidth for $|\Da| \leq 2
\pi \! \times \! 15$ GHz. Optical nonlinearities in this regime
may manifest as temporal correlations imposed on photons
transmitted through the cavity due to the light-induced atomic
vibrations, and will be the subject of future work.

%However, already for $\nbarnl < \omz / 2 \kappa \simeq 0.03$, the
%optical nonlinearity may become modified by the granularity of the
%optical forces upon the trapped atoms.  During its typical
%residence time of $1 / 2 \kappa$ in the cavity, a single photon
%imparts an impulse of $f(z) / 2 \kappa$ to an atom at location
%$z$. This impulse induces transient motional oscillations in the
%atomic medium that modulate the cavity resonance frequency. For
%our experimental parameters, with $N = 5 \times 10^4$ atoms, this
%modulation becomes greater than the cavity half-linewidth for
%$|\Da| \leq 2 \pi \! \times \! 15$ GHz. Optical nonlinearities in
%this regime could manifest as temporal correlations imposed on
%photons transmitted through the cavity due to the light-induced
%atomic vibrations. Such photon correlations will be the subject of
%future work.

Transient oscillations may also be excited coherently by
diabatically varying the probe-induced optical force. We observed
such oscillations as follows. With $\omz = 2 \pi \! \times \! 49$
kHz, we first loaded $N
> 5 \times 10^4$ atoms into the cavity giving $|\Delta_N| > 2 \pi \!
\times \! 20$ MHz for the atom-cavity detuning of $\Da = - 2 \pi
\! \times \! 260$ GHz, and then monitored the transmission of
probe light at $\Dc = -2 \pi \! \times \! 17$ MHz. As atoms were
lost from the trap, the atom number $N$ diminished until the
cavity resonance came into coincidence with the probe frequency
(Fig.\ \ref{fig:ring}(a)). Once the transmission signal reached a
threshold level, establishing conditionally the value of $\Delta_N
= -2 \pi \! \times \! 19$ MHz, the probe was switched off for
$10\,$ms and then turned back on within $5\,\mu$s at a high level
of $\bar{n}=6.5$. The potential minima within the combined
intracavity optical potential were thus suddenly displaced,
setting trapped atoms aquiver and causing the cavity resonance to
be modulated over $2 \beta = 0.75$ cavity half-linewidths at the
atom trapping frequency of $2 \pi \! \times \! 49$ kHz. The
corresponding oscillation amplitude (peak-to-peak) for all atoms
confined within a single representative well at $k_pz_0=\pi/4$
(Fig. \ref{fig:setup}(a)) is 0.8$\,$nm. For the Voigt profile
relevant to our system, the modulation of the cavity resonance
frequency varied the average cavity photon number by 1.3.  This
variation was visible in the temporal Fourier spectrum of the
cavity transmission signal from single experimental runs (Fig.\
\ref{fig:ring}(b), inset). By summing the transmission signals
over 50 repeated measurements, a clear temporal variation of the
cavity transmission, representing a record of the collective
motion of atoms within the cavity, was obtained.  The amplitude
and phase of this motion agreed with the predictions discussed
above. The observed motion decayed within $1 / \gamosc = 1\,$ms, a
value consistent with the inhomogeneous broadening of $\omz$ due
to the finite radial extent of the trapped gas.

For the present system, the observed damping of the collective
atomic motion imposes a technical limit of $\nb = \gamosc /2
\kappa \simeq 10^{-4}$ on the lowest light level at which these
motion-induced nonlinearities may be exploited. However, this
damping can be mitigated by radially confining the atoms more
tightly or cooling them to lower temperatures. Longer-lived
coherence may also be attained upon Bose-Einstein condensation,
which would occur for our system below an atomic temperature of
$T_c \leq 0.5\, \mu$K, close to that achieved experimentally.

%The use of ultracold atomic gases introduces new characteristic
%frequencies related to the atomic motion, $\omz$ and $\gamosc$
%herein, and thereby new physical regimes to cavity QED.

The present work highlights novel capabilities enabled by the
long-lived coherence of atomic motion inside an optical cavity.
Specifically, we observe strongly nonlinear optics, such as cavity
bistability, occurring at extremely low light levels. The role of
such nonlinearities in producing nonclassical and correlated quantum
states of light, such as achieved by saturation in single-atom
cavity QED \cite{turc95phase,duan04scale,birn05block}, warrants
further investigation.  Further, the influence of collective atomic
motion over cavity properties may allow for quantum-limited
measurement of that motion and for studies of quantum feedback
\cite{stec04qfb,stec06qfb,vule07external}.

We thank T.\ Purdy and S.\ Schmid for assistance in the early stages
of this experiment and H.J.\ Kimble for helpful comments. This work
was supported by AFOSR, DARPA, and the David and Lucile Packard
Foundation.

\bibliographystyle{apsrev}
%\bibliography{allrefs}

\begin{thebibliography}{29}
\bibitem{intnote}
\bibinfo{note}{For the large $\Da$ explored in this work, the equivalent
  nonlinearity parameter due to atomic saturation is $\beta \simeq
  (\DN/\kappa)(g_0^2/\Da^2)\nb$. For the parameters of Fig 2(a), this would
  imply $\nb\!>\!10^{4}$ photons for significant nonlinearity.}
\end{thebibliography}

\end{document}